\documentclass{article}

\usepackage{microtype}
\usepackage{graphicx}
\usepackage{subcaption}
\usepackage{booktabs}
\usepackage{placeins}

\newcommand{\evo}{Evo\,2}
\newcommand{\layer}{layer-26}

\usepackage{hyperref}

\usepackage[preprint]{icml2025_preprint}

\usepackage{amsmath}
\usepackage{amssymb}
\usepackage{mathtools}
\usepackage{amsthm}
\usepackage{listings}

\usepackage[capitalize,noabbrev]{cleveref}
\crefname{appendix}{Appendix}{Appendices}
\Crefname{appendix}{Appendix}{Appendices}

\usepackage{xcolor}

\newcommand{\revisioncolor}{black}
\DeclareRobustCommand{\rev}[1]{\textcolor{\revisioncolor}{#1}}

\theoremstyle{plain}

\theoremstyle{definition}

\theoremstyle{remark}

\icmltitlerunning{Screening of biosecurity features in metagenomic data with \evo{} probes}

\begin{document}

\twocolumn[
\icmltitle{Screening of Biosecurity Features in Metagenomic Data with Evo 2 Probes}

\icmlsetsymbol{equal}{*}

\begin{icmlauthorlist}
\icmlauthor{Jeremy Guntoro}{equal,jic,hackathon}
\icmlauthor{Alexander Dack}{equal,imperial,hackathon}
\icmlauthor{Dylan Danno}{hackathon,is}
\icmlauthor{Michaela Jančovičová}{imperial,hackathon}
\icmlauthor{Križan Jurinović}{imperial,hackathon}
\icmlauthor{Vanessa Smilansky}{hackathon,is}
\end{icmlauthorlist}

\icmlaffiliation{hackathon}{AIxBio Hackathon contestant}
\icmlaffiliation{imperial}{Department of Bioengineering, Imperial College London}
\icmlaffiliation{jic}{John Innes Centre}
\icmlaffiliation{is}{Independent Research Scientist}

\icmlcorrespondingauthor{Alexander Dack}{ad1114@ic.ac.uk}
\icmlcorrespondingauthor{Jeremy Guntoro}{jeremy.guntoro@jic.ac.uk}

\vskip 0.3in

\begin{abstract}
Genomic foundation models such as Evo~2 learn rich sequence representations, but
their value for biosecurity screening is largely unexplored. We ask how much
biosecurity-relevant signal is linearly accessible in these representations by
training minimal linear and attention probes on frozen Evo~2 layer-26
activations, without fine-tuning the underlying model. Across held-out
metagenomic test sets, the probes detect antimicrobial resistance (AMR) with
\rev{strong discrimination}: a linear probe reaches a region-level ROC-AUC of 0.888
(mean-pool), rising to 0.977 with a single-head attention probe. The probes
resolve finer-grained AMR drug-class subcategories and separate them from
unrelated functional genes, \rev{providing additional evidence that the learned
signal is not explained solely by generic functional-gene status}. Bacterial
virulence is also decodable, though more weakly (region-level ROC-AUC 0.833).
\rev{The AMR probe also retains comparable ranking performance on simulated
short reads without retraining, enabling evaluation before assembly in settings
where assembly is computationally costly or unreliable, such as low-coverage or
strain-heterogeneous communities. It achieves a read-level ROC-AUC of 0.898
(mean-pool), comparable to the mean-pooled full-region result. Within SynGenome,
AMR-associated prompt labels are only weakly recoverable from Evo~1.5-generated
sequences; these prompt-derived labels do not establish the function of the
generated response sequences.} A complementary sparse-autoencoder
analysis recovers interpretable resistance-associated features but proves less
consistent than the supervised probes. Together these results position
lightweight embedding-based probes as a fast, inexpensive first-pass detection
layer for metagenomic biosurveillance, and map
both the strengths and the current limits of the approach. This work was conducted as part of the \textit{AIxBio Hackathon 2026} hosted by \textit{BlueDot Impact}, \textit{Apart Research}, and \textit{Cambridge Biosecurity Hub}.
\end{abstract}
]
\icmlkeywords{Machine Learning, Genomics, Biosecurity, Interpretability}
\printAffiliationsAndNotice{\icmlEqualContribution}
% =====================================================================
\section{Introduction}
\label{sec:introduction}
% =====================================================================

Advances in genomic foundation models have enabled rich embedding representations of biological sequences, yet their application to biosecurity-relevant tasks remains limited. In this work, we investigate the use of Evo 2-derived embeddings combined with probing and sparse autoencoder (SAE) techniques for antimicrobial resistance (AMR) detection in metagenomic data. We demonstrate that embedding-based representations enable accurate identification of AMR signals from both metagenome-assembled genome (MAG) derived sequences as well as simulated metagenomic reads across diverse, multi-organism datasets. For MAG-derived sequence regions, our approach captures not only the presence of AMR genes but also finer-grained subcategories of resistance, as well as bacterial virulence factors, indicating that these properties are encoded within the representation space. At the simulated read level, we show that embeddings can recover AMR signals from fragmented and heterogeneous sequencing data, highlighting their potential for real-time surveillance in complex environments. Our results suggest that genomic embeddings provide a scalable and generalizable framework for resistome profiling, and that interpretability methods such as probing and SAEs offer a path toward mechanistic insight into learned biological features. This work establishes a foundation for embedding-based biosecurity applications, including the detection of emerging or engineered resistance elements in metagenomic data.

\section{Related work}
\subsection{Antimicrobial Resistance Detection from Metagenomic Sequencing}
Antimicrobial resistance (AMR) detection from metagenomic sequencing has traditionally relied on sequence homology against curated databases such as CARD, ResFinder, and AMRFinderPlus, using alignment of reads or assembled contigs to reference resistance gene collections \cite{alcock_card_2019,florensa_resfinder_2022,feldgarden_amrfinderplus_2021}. 

More recent approaches include genome-scale foundation models and deep learning methods for metagenomic sequence representation and antimicrobial resistance prediction. METAGENE-1 is a 7B parameter autoregressive transformer trained on approximately 1.5 trillion base pairs of wastewater-derived metagenomic sequences, producing general-purpose sequence representations evaluated on pathogen detection and other genomic classification and embedding-based benchmarks \cite{liu_metagene-1:_2025}. In parallel, DeepARG applies deep neural networks trained on known antibiotic resistance gene families to classify metagenomic sequences, enabling prediction of resistance-associated labels directly from sequence data~\cite{arango-argoty_deeparg:_2018}.

SecureBio’s Outward Assembly pipeline is used in wastewater metagenomic sequencing workflows to iteratively extend contigs from short seed sequences flagged by upstream detection methods, enabling reconstruction of local genomic context around low-abundance genomic signals \cite{fields2025outwardassembly}. This type of escalation-based reconstruction is complementary to embedding-based approaches and may provide useful downstream context for interpreting signals identified by representation learning methods such as those explored in this work.

\subsection{Genome Foundation Models}
A range of genomic foundation models have been proposed for learning representations from DNA sequences, including DNABERT~\cite{ji_dnabert:_2021}, the Nucleotide Transformer~\cite{dalla-torre_nucleotide_2025}, HyenaDNA~\cite{nguyen_hyenadna:_2023}, METAGENE-1~\cite{liu_metagene-1:_2025}, EVO~\cite{nguyen_sequence_2024}, and Evo~2~\cite{brixi2026evo2}. Evo~2 is used in this work as it represents a recent large-scale, openly available genome foundation model trained across the full tree of life, and provides accessible intermediate representations suitable for downstream analysis.

\subsection{Interpreting Genome Foundation Models}
Understanding what genomic foundation models encode is an active area of research. A standard approach is to treat hidden states as fixed embeddings and evaluate them using linear probes, a paradigm widely used across genomic and protein language models. These studies show that pretrained representations contain biologically meaningful signals that are often linearly accessible, particularly in intermediate layers.

More expressive interpretability methods such as sparse autoencoders (SAEs) have also been applied to both protein and genomic models, recovering structured biological features including regulatory elements, coding structure, and protein-level annotations, indicating that learned representations encode compositional biological structure across domains.

Within this setting, Evo~2 has become a key reference model due to its scale and open availability. Prior work has shown that Evo~2 embeddings support strong downstream prediction via simple probes (e.g. EVEE for variant effect prediction~\cite{pearce_evee:_2026}) and that SAEs can recover interpretable biological features and enable feature-based analysis (e.g. Goodfire-style analyses~\cite{goodfire2025interpreting}), reinforcing that meaningful biological structure is present in its latent space.

Our work builds on this line of research by applying linear probes to Evo~2 hidden representations for antimicrobial resistance prediction, extending prior probing and embedding-based studies from regulatory and variant-effect settings to AMR screening in metagenomic sequences. The most closely related recent work is by \citet{tai_cross-species_2026}, which
investigates cross-species AMR prediction using Evo-based genomic foundation
model embeddings. They classify whole genomes by resistance phenotype and
mitigate species-composition confounds through locality-preserving aggregation,
whereas we detect resistance genes at the level of individual coding sequences
and control the same confound directly through matched negatives.

% =====================================================================
\section{Methods}
\label{sec:methods}
% =====================================================================

\subsection{\evo{}: A DNA Foundation Model}
\label{sec:evo}

Our analysis focuses on \evo{}-7B-262k \cite{brixi2026evo2}, a 7-billion-parameter
DNA foundation model that accepts
as input a string of DNA and outputs next-token predictions. However, an
alternative use case for such foundation models is to treat the model not as a
generator but as a feature extractor: rather than reading off the output
distribution, we run a forward pass over an input sequence and retain the
internal residual-stream activations at a chosen layer. For a sequence of $n$
tokens, layer $\ell$ produces an activation matrix of shape $n \times d$, where
$d=4096$ is the model's hidden dimension, hence each row is a $4096$-dimensional
embedding of one token in its sequence context. Specifically, we extract
activations from \layer{} to train linear and attention probes
(\cref{sec:training}), as well as to map feature activations using the Goodfire
Sparse Autoencoder \cite{brixi2026evo2,goodfire2025interpreting} (\cref{sec:sae}). Layer-26 was
initially selected as a target because existing work on the Goodfire Sparse
Autoencoder has previously demonstrated activation of interpretable features in
response to fundamental biological elements, such as the presence of coding
sequence (CDS) or promoter regions \cite{brixi2026evo2,goodfire2025interpreting}. We therefore
hypothesised that this layer may also encode biosafety-relevant features. Full
implementation and infrastructure details are given in \cref{app:implementation}.

\subsection{Datasets}
\label{sec:data}

We utilize three datasets. The first is MGnify \cite{richardson2023mgnify}, a database of metagenome-assembled genomes
(MAGs), from which we extract coding sequences together with their flanking
genomic context from the chicken gut (816 MAGs used) and human skin (255 MAGs used) catalogues, spread over 901 distinct species of bacteria. The second is the Virulence Factor Database (VFDB) \cite{liu2022vfdb}, from which we extract virulence factor CDS over 34 species. Only CDS from species with a representative MAG from the MGnify dataset were used, as negative training examples were drawn from MGnify. The third is SynGenome
\cite{merchant2025semantic}, a database of Evo-1.5-generated sequences up to 5\,kb in
length, each containing a coding sequence. Each of these datasets contains sets of labels for each sequence. Our aim is to train probes that can identify the presence or
absence of specific labels based on the \evo{} \layer{} embeddings of their
underlying sequences. Specifically, we train probes to detect the following:
bacterial antimicrobial resistance (AMR, also evaluated on simulated short reads), AMR resolved by individual drug class, bacterial virulence (BV), \rev{and recovery of AMR-associated prompt labels from SynGenome-generated sequences}.

Full details of these datasets, along with the labels they contain, may be found
in \cref{tab:datasets}.

\begin{table*}[!htbp]
  \centering
  \caption{Datasets used to train and evaluate the probes. \rev{The first four rows define three detection tasks: binary AMR, bacterial virulence (whose positives and negatives are listed separately), and the SynGenome prompt-label task.} BV negatives are drawn from the MGnify sets (matched by species). ``Split unit'' is the grouping kept intact across the split: no MAG or species is shared between train, val and test, so evaluation is
  always on unseen MAGs (for AMR) or species (for BV); \rev{SynGenome is split at the individual-record level} (\cref{sec:training}). The last two (MGnify STRESS and VIRULENCE) are
    used only as transfer-negative evaluation sets for the per-drug-class AMR probes
    (\cref{tab:perclass}); no probe is trained on them, and their counts are the
    held-out test-split pools used in evaluation. The ``Extent'' column
    gives what each record spans: MGnify AMR is a CDS with $\pm$2\,kb genomic
    flank; VFDB is the CDS alone; SynGenome is a fixed $\leq$5\,kb
    Evo~1.5-generated
    window with an embedded ORF but no CDS coordinates (so the whole window is
    labelled).
    }
  \label{tab:datasets}
  \begin{tabular}{l l r l l}
    \toprule
    Dataset & Label / content & Records (pos $+$ neg) & Split unit & Extent \\
    \midrule
    MGnify MAG (AMR) & AMR CDS & 4{,}458 (2{,}229 $+$ 2{,}229) & MAG & CDS $\pm$2\,kb flank \\
    VFDB BV positives & BV CDS & 7{,}325 positives only & Species & CDS only \\
    MGnify BV negatives & non-BV CDS & 7{,}325 negatives only & Species & CDS only \\
    \rev{SynGenome generated sequences} & \rev{AMR-associated prompt label} & 4{,}458 (2{,}229 $+$ 2{,}229) & \rev{Record} & $\leq$5\,kb window \\
    \midrule
    MGnify MAG (STRESS) & stress-response CDS & 47 & --- & eval-only \\
    MGnify MAG (VIRULENCE) & bacterial virulence CDS & 31 & --- & eval-only \\
    \bottomrule
  \end{tabular}
\end{table*}
\subsection{Training Linear and Attention Probes}
\label{sec:training}

We trained minimal single-layer probes that accept as input the per-token
(i.e., per-nucleotide position) \layer{} activations, where each nucleotide is
represented by a 4096-dimensional vector, so a sequence of length $n$ has input
dimension $n\times 4096$. Each probe maps this variable-length activation
sequence to a single per-region logit. We tested both linear and attention probes
\cite{alain2017probing,kantamneni2025attention,mckenzie2025highstakes} for AMR,
but only linear architectures for the remaining labels.

Linear and attention probes both assign each nucleotide a scalar
\emph{per-token logit} $z_i$ via a learned linear map of its activation, and
aggregate these into the per-region logit as a weighted sum
$s=\sum_{i=1}^{n}\alpha_i z_i$ with $\sum_i\alpha_i=1$. \rev{At inference they differ in how the weights $\alpha_i$ are set, while their training supervision also differs as described below.} The linear probe uses uniform weights,
$\alpha_i=1/n$, so the region logit reduces to the mean of the per-token logits (mean-pooling),
$s_{\text{mean}}=\frac{1}{n}\sum_i z_i$. The attention probe instead uses input-dependent \emph{attention weights} for $\alpha_i$, which are computed by softmax attention from a learned query, so that nucleotides that are more informative for classification receive larger weightings (see Appendix \ref{app:implementation} for further details).

The input to each probe is a genomic region encoded by \evo{}. What constitutes
a region depends on the dataset and is summarised in \cref{tab:datasets}: in
some cases it is a coding sequence on its own, and in others a coding sequence
together with flanking genomic context. For the AMR data, for example, each
region is a coding sequence extended by up to 2\,kb of flanking sequence on
either side, truncated where the contig is too short to accommodate the full
flank. For VFDB the region is the coding sequence alone, and for SynGenome the region is a CDS-containing window. In every case the probe receives the per-token \layer{}
activations of the region. Linear probes are trained with per-token supervision, where only tokens inside the labelled CDS are marked positive. Attention probes, by contrast, are trained with a single label over the whole region, since their architecture pools across tokens into one region-level prediction.

For probes, a common failure mode is that they learn to
detect features distinct from the feature of interest
\cite{kantamneni2025attention,mckenzie2025highstakes}, and hence care was taken to control for other
features. For the binary AMR probe, each positively labelled region was paired 1:1 with a negative region drawn from the same MAG. For the bacterial virulence probe, each virulence-factor CDS was paired against a non-virulence-factor CDS from the same species from the MGnify database (see \cref{tab:datasets}). \rev{These split units were chosen to prevent the same MAG from crossing the AMR splits and the same species from crossing the bacterial-virulence splits, thereby reducing contextual and taxonomic leakage.} Regions are then matched in CDS length to within
$\pm$20\% and in GC content to within $\pm$0.05 absolute of the paired positive. Negative CDSs were additionally required not to overlap any
AMR/STRESS/VIRULENCE hit, antiSMASH biosynthetic gene cluster, CRISPR array,
defence-system region, or mobilome element. The overwhelming majority of positives (99.6\% for AMR, 94.7\% for VFDB) were matched under these strict criteria. A fallback hierarchy, given in \cref{tab:negmatch} of \cref{app:implementation}, was applied for the remaining data. \rev{Unpaired positives were discarded, leaving 2{,}229 AMR and 7{,}325 VFDB positive--negative pairs in the analyses.} The per-drug-class AMR probes are trained differently, using other AMR drug classes as negatives, so this length/GC/MAG matching does not apply to their training set. On the other hand, the per-drug-class probes are still evaluated against the matched-negative pool (NEG), to check that they activate in response to the correct features. This pool is matched at the population level rather than per positive: it comprises every matched negative CDS available in the held-out MAGs — the 336 negatives paired to AMR CDS together with the 78 drawn for the STRESS (47) and VIRULENCE (31) anchor regions - and is not re-matched to each drug class. All 414 negatives pass the same exclusion criteria, and the two subpools are indistinguishable in CDS length and GC content (median 981 bp and GC 0.47 for both), so the broader composition does not materially affect the contrast. For the SynGenome set, no matching process was attempted.

The goal of training is to obtain a probe that will reliably fire only on embeddings of regions containing CDS that are associated with a specific label. Our primary measure of performance is therefore the ROC-AUC. For training we
apply a binary cross-entropy loss and the Adam optimiser~\cite{kingma_adam:_2017}. Datasets are split
into training, validation, and test sets. The best epoch is selected by maximising ROC-AUC on
the validation set. All reported ROC-AUC values are
evaluated on the held-out test set, which is disjoint from training and
validation. The reported F1 values (\cref{tab:results}) use an F1-maximising threshold selected on the validation split and then applied to the held-out test split. Per-probe dataset splits are given in
\cref{tab:splits}, and full training details in \cref{app:implementation}.
\FloatBarrier
\begin{table*}[htbp]
  \centering
  \caption{Region-level probe performance across the datasets. For the AMR attention probe, ROC-AUC is computed over token logits attention-weighted and summed over the whole region, while for the linear probes, per-token logits are mean-pooled over each region and ROC-AUC is computed on the pooled score. F1 is calculated at each probe's F1-maximising threshold, selected on the validation split and applied to the test split. For the short-read row the threshold is selected on 301 bp
  segments of the validation CDS, since region-level and read-level scores are not on a common scale.}
  \label{tab:results}
  \begin{tabular}{l l c c}
    \toprule
    Dataset & Probe & ROC-AUC & F1 \\
    \midrule
    AMR binary        & Linear          & 0.888 & 0.823 \\
    AMR binary        & Attention                & 0.977& 0.923 \\
    AMR binary short read & Linear &  0.898 & 0.809 \\
    Bacterial virulence (BV)      & Linear     & 0.833 & 0.774 \\
    \rev{SynGenome prompt-label task} & Linear & 0.599 & 0.663 \\
    \bottomrule
  \end{tabular}
\end{table*}

\subsection{Simulating Metagenomic Read-level Data}
\label{sec:reads}

To evaluate probe performance on realistic sequencing data, we simulated Illumina sequencing reads using the Python package InSilicoSeq \citep{gourle_simulating_2019} and evaluated the performance of the pre-trained linear probe on the resulting reads. InSilicoSeq generates realistic metagenomic sequencing data by modelling Illumina-specific sequencing characteristics, including substitution, insertion, and deletion errors, GC bias, and insert size distributions. We used the pre-built MiSeq error model, which produces paired-end reads with an average length of approximately \rev{301 bp}, closely matching the output of an Illumina MiSeq instrument. 

A balanced set of AMR-positive and AMR-negative coding sequences (CDSs) was constructed from multiple organisms to represent a heterogeneous metagenomic environment. Reads were then simulated directly from these CDS regions, allowing sequencing errors to be introduced according to the MiSeq error profile while preserving the underlying AMR labels. The resulting dataset comprised short, noisy sequencing reads originating from diverse organisms and was used to evaluate the robustness of the probe on realistic metagenomic sequencing data. A multi-organism evaluation set of 1{,}340 reads ($2$ reads $\times$ $335$ CDS $=$ $670$ AMR-positive samples
$+$$2$ reads $\times$ $335$ CDS $=$ $670$ AMR-negative samples; note $335$ CDS were used as one CDS was missing from the set of simulated reads, and its matched counterpart was excluded to preserve the paired, balanced evaluation set) was assembled. Each read is embedded and scored with the same
\layer{} pipeline and pooling rules as full regions (\cref{sec:evo,sec:training});
results appear in \cref{sec:amr-reads}. Note reads are embedded without their genomic context, so read and full-region scores are not on a common scale. \rev{The read F1 threshold is therefore selected on one randomly sampled clean 301\,bp segment per validation CDS, each scored by a fresh forward pass, and is then applied without further adjustment to the InSilicoSeq-generated test reads. This adapts the operating threshold to sequence length while retaining a distribution shift in sequencing noise.}

\subsection{Sparse Autoencoder Feature Analysis}
\label{sec:sae}
As an alternative to supervised probes, we also investigated sparse autoencoder (SAE)-based approaches for identifying genomes of biosecurity interest. This is motivated by prior work \cite{brixi2026evo2}, which demonstrated that SAE-derived features can capture generic genomic characteristics, including prophage regions in prokaryotic genomes.

Our work builds upon the sparse autoencoder released by the Goodfire and Arc Institute Evo-2 research team, available via \href{https://huggingface.co/Goodfire/Evo-2-Layer-26-Mixed}{Goodfire Evo-2 SAE}. The model takes the 4096-dimensional activations from layer-26 of the Evo 2 7B model and maps them to a 32,768-dimensional sparse latent representation.

To identify candidate latent dimensions associated with genomic annotations of interest, we applied the SAE to the token-level layer-26 activations of each input sequence. For each sequence and each SAE feature, activations were averaged across all tokens in the input region, including flanking genomic context where present. We then compared these region-level mean activations between annotation groups within the binary-AMR dataset.

This analysis highlights sparse dimensions that are consistently more strongly activated across feature-positive regions and may therefore encode feature-relevant genomic signal. For downstream analysis, we prioritised dimensions showing large and consistent separation between the two groups rather than applying a fixed numerical threshold. Because feature selection and visualisation were performed on the same dataset, the resulting dimensions should be interpreted as descriptive candidate associations rather than independently validated features.

% =====================================================================
\section{Results}
\label{sec:results}
% =====================================================================

\subsection{\evo{} probes}
\label{sec:evo-probes}

\subsubsection{Bacterial antimicrobial resistance (AMR)}
\label{sec:amr}

\begin{figure*}[htbp]
  \centering
  \includegraphics[width=0.752\textwidth]{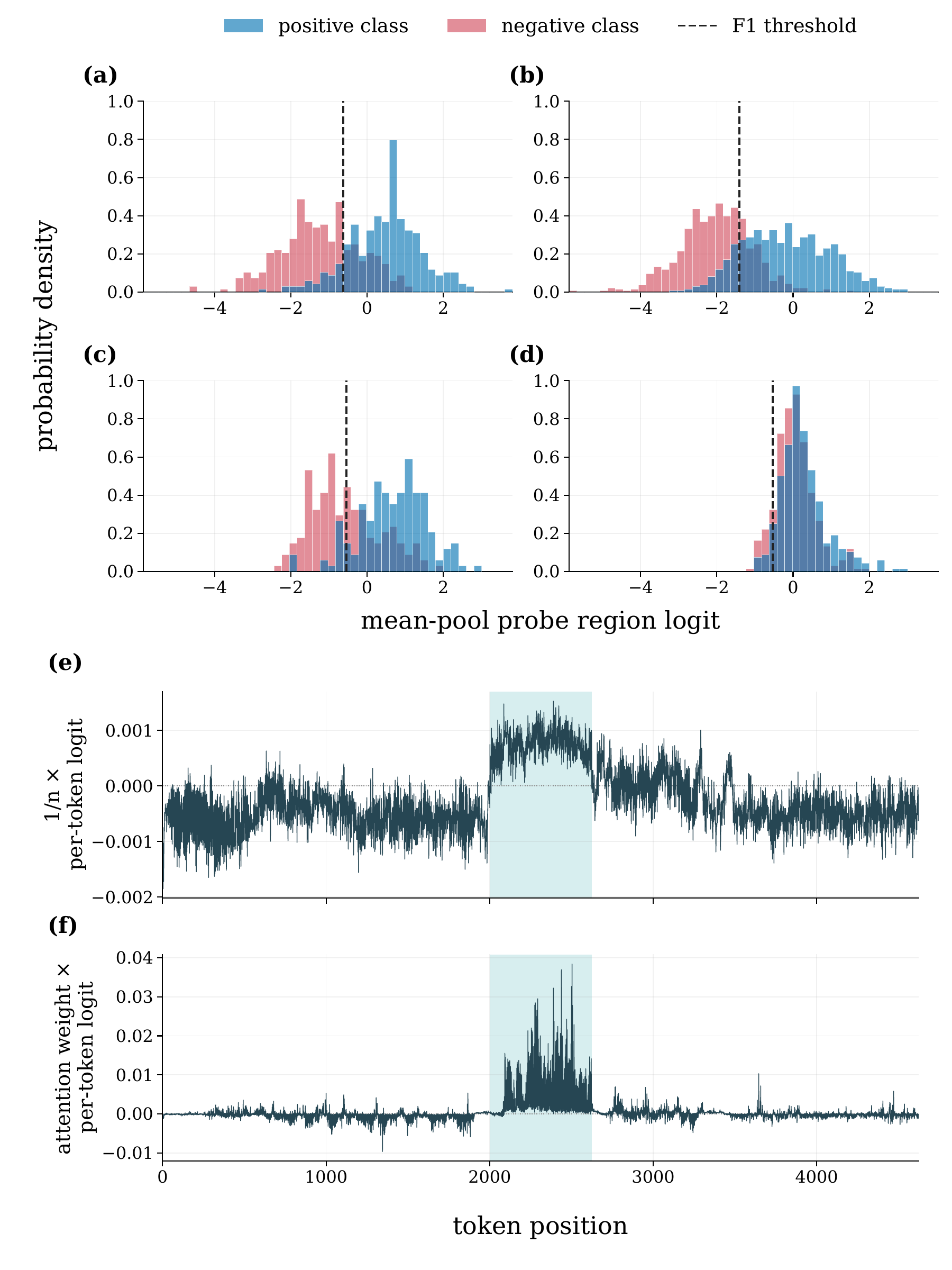}
  \vspace{0.0cm}
  \caption{Mean-pooled distributions of per-region linear probe scores across datasets. The linear probe produces a score at each token (i.e., nucleotide position), which is then mean-pooled over annotated genomic regions (including flanking regions where present and coding sequences) to obtain a single score per region; these region-level scores are aggregated into distributions over positive and negative regions. \rev{(a) AMR probe applied to CDS-centred regions extracted from metagenome-assembled genomes (MAGs), using CDS-level antimicrobial-resistance (AMR) annotations. (b) AMR probe evaluated on simulated Illumina short reads generated with a MiSeq sequencing-error model.} (c) VFDB bacterial virulence probe applied to coding sequence (CDS) regions only, without flanking context, using CDS-level bacterial virulence annotations. \rev{(d) SynGenome probe evaluated on AMR-associated versus non-AMR-associated prompt labels.} Positive regions are shown in blue and negative regions in pink. Dashed lines mark the F1-maximising threshold, selected on the validation split; for (b), on 301 bp segments of the validation CDS. (e,f) Per-token probe contributions across an assembled DNA region from \textit{Onthenecus intestinigallinarum} (locus MGYG000307601\_01495 from the MGnify chicken gut MAG database), with the chloramphenicol-resistance coding sequence (CDS) shaded. (e) Linear probe contributions, computed as the per-token logit scaled by $1/n$, where $n$ is the length of the DNA region. (f) Attention-based probe contributions, computed as attention weights multiplied by the per-token logit. \rev{For this example,} the contribution is concentrated within the annotated resistance gene in both cases, with the attention-based probe exhibiting sharper localisation and near-zero contribution outside the CDS.}
  \label{fig:hists}
  %\caption{Mean-pooled distributions of per-region linear probe scores across datasets. The linear probe produces a score at each token, which is then mean-pooled over tokens within annotated genomic regions (including flanking regions where present and coding sequences) to obtain a single score per region; these region-level scores are aggregated into distributions over positive and negative regions. (a) AMR probe applied at the metagenome-assembled genome (MAG) level, where genomes are labeled as antimicrobial resistance (AMR) positive or negative based on genome-level annotations. (b) AMR probe evaluated on simulated Illumina short reads generated to mimic real sequencing noise and environmental sampling conditions. (c) VFDB bacterial virulence probe applied to coding sequence (CDS) regions only, without flanking context, using CDS-level bacterial virulence annotations. (d) Generative AMR probe evaluated on SynGenome sequences. Positive regions are shown in blue and negative regions in pink. Dashed lines mark the F1-maximising threshold.}
\end{figure*}

\paragraph{Linear and attention probes.}
Both probes achieve high region-level ROC-AUC (\cref{tab:results}). \rev{The attention probe, which is trained with a region-level objective and pools tokens via learned attention into a single region logit, achieves a higher region-level ROC-AUC than the token-supervised linear probe.} We set each probe's decision threshold to its F1-maximising value on the validation split. Score distributions
for positive and negative regions are shown in \cref{fig:hists}(a)--(d). \rev{In the example shown in \cref{fig:hists}(e)--(f), both probes concentrate their contribution within the annotated CDS, the expected behaviour if the intended label is being detected. The attention probe also shows this CDS-specific response despite being trained on whole-region labels.}

%\FloatBarrier
%\begin{figure}[htbp]
%  \centering
%  \includegraphics[width=0.99\columnwidth]{fig_contrib}
%  \caption{ Weighted per-token probe contributions across a DNA region from \textit{Onthenecus intestinigallinarum} (locus MGYG000307601\_01495 from the MGnify chicken gut MAG database), with the chloramphenicol-resistance CDS shaded. (a) Linear probe and (b) attention-based probe, as defined in Section~\ref{sec:training}. In both cases, ``per-token'' refers to per-nucleotide positions. For the linear probe (a), contributions are given by the per-token logit scaled by $1/n$, where $n$ is the length of the region, applied to the logits at each nucleotide position. For the attention-based probe (b), contributions are given by attention weights multiplied by the per-token logit, with weights varying across nucleotide positions. In both cases, signal concentrates within the CDS (coding sequence), consistent with the model detecting the intended label; the attention-based probe shows sharper localisation with near-zero contribution outside the CDS.}
%  \label{fig:amr-contig}
%\end{figure}

\begin{table*}[htbp]
  \centering
  \caption{Per-drug-class AMR probes (one-vs-rest within AMR). In-distribution
    ROC-AUC separates each class from the other AMR classes; the transfer
    columns instead use NEG, STRESS, and VIRULENCE negatives. NEG is the matched
    non-AMR negative set of the MGnify MAG AMR dataset (extended with the negatives that the matching pipeline drew for the STRESS and VIRULENCE regions), and STRESS and
    VIRULENCE are functional categories used as alternative transfer negatives.}
  \label{tab:perclass}
  \begin{tabular}{l c c c c}
    \toprule
    Drug class & In-dist.\ & vs.\ NEG & vs.\ STRESS & vs.\ VIRULENCE \\
    \midrule
    Glycopeptide  & 0.996 & 0.954 & 0.987 & 0.990 \\
    Macrolide     & 0.994 & 0.958 & 0.970 & 0.983 \\
    Beta-lactam   & 0.998 & 0.963 & 0.962 & 0.783 \\
    Aminoglycoside& 0.989 & 0.967 & 0.979 & 0.972 \\
    Phenicol      & 0.992 & 0.976 & 0.958 & 0.986 \\
    \bottomrule
  \end{tabular}
\end{table*}

\paragraph{Per-drug-class AMR probes.}
We trained five separate linear probes, one for each of the five most common
AMR drug classes (glycopeptide, macrolide, beta-lactam, aminoglycoside, and
phenicol), in a one-vs-rest setting: for a given class, positives are in-CDS
tokens of that class and negatives are in-CDS tokens of the other four AMR
classes. Each probe is then evaluated on four streams on the held-out test split
(\cref{tab:perclass}; dataset splits in \cref{tab:splits-perclass}). The in-distribution stream is the one-vs-rest task above,
measuring how well the probe separates its own class from other AMR classes. The
three transfer streams instead evaluate the same trained probe against negatives it never saw in training, drawn from a different category: NEG  (the matched negative pool from the binary AMR probe, extended with the negatives that the matching pipeline drew for the STRESS and VIRULENCE regions), STRESS, and VIRULENCE, measuring
whether a class-specific probe also rejects unrelated functional genes. All
classes achieve in-distribution AUC\,$>$\,0.98, and transfer AUC generally
remains high ($>$\,0.95), \rev{providing further evidence that the signals are not explained solely by generic functional-gene status}. The one exception is the beta-lactam probe,
which scores lower against bacterial virulence (0.783); \rev{this weaker separation may reflect biological overlap, dataset composition, or the limited size of the virulence transfer set}. The STRESS and
VIRULENCE transfer pools are small (47 and 31 records; \cref{tab:splits-perclass}),
so these transfer AUCs, and the beta-lactam-versus-virulence value in particular,
carry correspondingly wide uncertainty.

\paragraph{AMR probes on simulated reads.}
\label{sec:amr-reads}
\rev{Applying the unchanged AMR binary-probe weights to simulated short reads (\cref{sec:reads}) retains comparable ranking performance without retraining (\cref{tab:results}): the read-level ROC-AUC of 0.898 is comparable to the mean-pooled full-region result of 0.888. After length-matched threshold calibration on clean validation fragments, the probe achieves an F1 of 0.809 on noisy simulated test reads.} Score distributions for positive and negative reads are shown in
\cref{fig:hists}(b).

\subsubsection{Bacterial virulence (BV)}
\label{sec:bv}

Bacterial virulence is decodable but harder than AMR (\cref{tab:results}). Score distributions for positive and negative
regions are shown in \cref{fig:hists}(c).

\subsubsection{\texorpdfstring{\rev{SynGenome AMR-associated prompt labels}}{SynGenome AMR-associated prompt labels}}
\label{sec:gen-amr}

We test whether AMR-associated prompt labels are recoverable from Evo 1.5-generated sequences within the SynGenome dataset \cite{merchant2025semantic}. A linear probe trained and tested entirely within
the SynGenome distribution reaches only mean-pool AUC
0.599 (\cref{tab:results}). Even in-distribution, \rev{AMR-associated versus non-AMR-associated SynGenome prompt labels are} weakly separable, with the positive and negative score distributions heavily
overlapping (\cref{fig:hists}(d)). Hence, under this protocol, the prompt-derived AMR labels are only weakly predictable from the generated response sequences. \rev{The validation-selected test F1 of 0.663 also does not exceed the 0.667 F1 obtained by predicting every example as positive on this balanced test set.}

\subsubsection{Sparse autoencoder features}
\label{sec:sae-results}

\begin{figure}[htbp]
\centering
\includegraphics[width=0.99\linewidth]{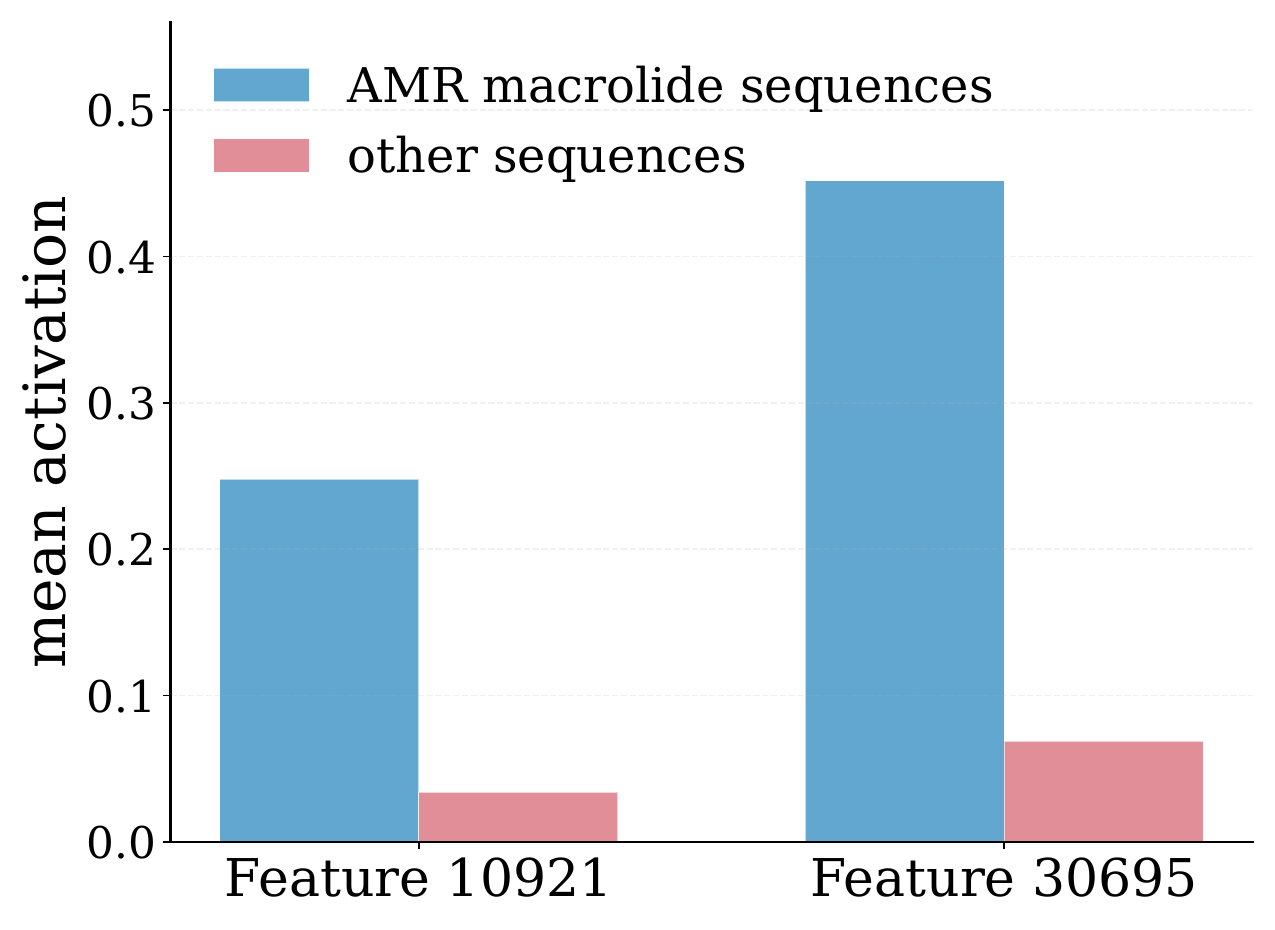}
\caption{ Exploratory comparison of sparse autoencoder features associated with macrolide antibiotic resistance in the AMR-binary dataset. We use a pre-trained SAE from Goodfire \cite{brixi2026evo2,goodfire2025interpreting}. Two salient features (SAE latent dimension) are plotted on the x-axis, with two bars per feature corresponding to different subsets of genomic regions. The SAE is applied to token-level embeddings, and values are aggregated by taking the mean over tokens within sequence regions. The y-axis therefore represents the mean SAE feature activation over the relevant region. Blue bars correspond to coding regions annotated as macrolide-associated antimicrobial resistance (AMR) genes, while red bars correspond to AMR-associated coding regions not linked to macrolide resistance. This comparison highlights features that preferentially activate for macrolide-specific resistance signals versus other AMR-related signals.}
\label{fig:macrolide_sae}
\end{figure}

We also examined whether individual features of the Goodfire \layer{} sparse autoencoder align with biosecurity-relevant annotations (\cref{sec:sae}). Using this approach, we identify several latent dimensions associated with antimicrobial resistance (AMR)-related signals. For example, dimensions 10921 and 30695 show consistently elevated activation for sequences associated with macrolide resistance across both the AMR-binary and AMR-binary-short-read settings (\cref{fig:macrolide_sae}).

On short-read DNA sequences, we observe a substantial reconstruction gap in the SAE. The L2 reconstruction error between input embeddings and reconstructed embeddings is approximately 25\% of the error between the input embeddings and a zero vector baseline. This indicates that while the SAE retains some signal, a non-trivial portion of the input representation is not preserved under the learned sparse encoding.
\FloatBarrier

\section{Discussion}
\label{sec:discussion}

Our central result is that a minimal probe on \evo{} \layer{} representations
detects a broad range of biosecurity-relevant features with \rev{strong discrimination},
spanning antimicrobial resistance, its finer-grained drug-class subcategories,
and bacterial virulence, despite using only a single linear or attention-pooling
layer over frozen activations. That so much signal is linearly accessible at a
single layer is encouraging: it suggests \evo{} encodes these functional
properties in a form that is cheap to read out, without task-specific
fine-tuning of the underlying model. Because each probe reads
directly from frozen activations, probes for many different properties can be
trained and evaluated in parallel at relatively small additional cost.
 
The open question is how this translates to a realistic metagenomic setting,
where sequences are fragmented into reads that are taxonomically heterogeneous and of variable
quality. It is especially encouraging that the probes, which were trained on long sequences, transfer well to our
simulated short-read data. Read-length fragments are the regime most relevant to
real metagenomic surveillance, and the probes retain strong performance on them
despite being trained on longer assembled regions. That said, simulated reads
are an optimistic proxy, so
establishing the probe's accuracy and sensitivity on genuine samples remains the
necessary next step.
 
\rev{Within SynGenome, the probe only weakly separates generated windows grouped by AMR-associated versus non-AMR-associated prompt labels (\cref{fig:hists}(d)). This is not a direct test of whether the generated sequences functionally confer AMR: the labels derive from the natural prompt context and do not establish that the generated response ORF has the corresponding function. The low separability could therefore reflect weak retention of prompt-associated signal, imperfect correspondence between prompt labels and generated ORFs, limitations of the probe, or some combination of these factors. Wet-lab validation would be required to determine which generated sequences are genuinely functional. This caution is consistent with \citet{ikonomova2025experimental}, who found that AI-generated sequences satisfying \textit{in silico} structural constraints did not reliably retain the intended biological function. Their results concern protein redesign with models outside the \evo{} lineage and therefore do not transfer directly to SynGenome. Conversely, \citet{king2025generative} showed that closely related Evo-family models can generate experimentally viable bacteriophage genomes, although only 16 of roughly 300 tested designs were viable and generation was tightly guided by a natural template. Thus, experimental activity cannot be inferred from a prompt-associated label alone.}
 
Overall, while SAE-based approaches showed some promising signals, probe-based
methods ultimately demonstrated more consistent performance within
this work. This comparison was not like-for-like, however: our probes were
trained directly on the target data distributions, whereas the sparse
autoencoder was pre-trained and applied out of distribution. Because SAEs are
known to be sensitive to distributional shift, weaker performance from an
off-the-shelf SAE is expected. Even so, the SAE still appeared to encode AMR-relevant structure, with
consistent activation patterns across datasets, suggesting that an SAE trained
directly on this domain could perform considerably better. Establishing that
fairly will require both a domain-trained SAE and larger, more diverse
evaluation sets.
 
Several directions follow directly from these results. The most important is
evaluation against a lab-verified synthetic-genome dataset: pairing generated
sequences with experimental confirmation of function would \rev{link prompt-associated labels to verified properties of the generated sequences and turn the present ambiguity into a quantitative benchmark}. The same probing approach could also be extended to monitoring
engineered synthetic-biology parts \citep{igem_registry}, where the
ability to flag hazardous functional elements such as virulence
factors, toxins, or deliberately introduced resistance determinants in
assembled or fragmented constructs would have direct biosecurity
value, for instance in screening synthetic-DNA orders or detecting engineered
pathogens. Methodologically, the
attention probe was applied only to the AMR binary task; all other results use
linear probes alone. Since the attention probe outperformed the linear probe on AMR, extending it to the bacterial virulence, per-drug-class, and read-level tasks is likely to
improve those results as well. The reconstruction degradation we
observe on short-read sequences---consistent with a mismatch between the
pre-trained SAE's training distribution and our shorter reads---also motivates
training a sparse autoencoder directly on short-read embeddings, which could
improve both reconstruction fidelity and interpretability in the metagenomic
regime. Then, validation on real metagenomic datasets, as
opposed to MAG-derived or simulated reads, would test the probes under genuine
deployment conditions and quantify the accuracy and sensitivity that our
discussion of the metagenomic setting leaves open. Finally, the same
approach extends naturally to \rev{metatranscriptomic cDNA sequencing}, where
reverse-transcribed reads would let biosecurity-relevant features be probed in
RNA viruses as well as DNA organisms; a complementary priority is to
characterise the computational cost of scaling \evo{}-embedding extraction to
high-throughput metagenomic read volumes, since this will set the throughput
achievable in a deployed first-pass screen.
 
\section{Conclusion}
\label{sec:conclusion}
 
We have shown that minimal probes on the intermediate representations of \evo{}
can detect a range of biosecurity-relevant signals, including antimicrobial
resistance and its drug-class subcategories as well as bacterial virulence,
directly from genomic sequence and without fine-tuning the underlying model.
These signals are largely linearly accessible at a single layer. \rev{The region-supervised attention probe achieved higher region-level discrimination than the token-supervised mean-pooled probe, although this comparison does not isolate attention from the difference in training objective. The unchanged linear-probe weights retained comparable ranking performance on simulated short reads. Within SynGenome, AMR-associated prompt labels were only weakly recoverable from generated sequences; this result does not directly measure functional AMR in the generated ORFs.} 
An exploratory sparse-autoencoder analysis identified candidate features associated with macrolide-labelled regions, although these associations were not independently validated and reconstruction degraded on short-read inputs. Taken together, these findings suggest that
embedding-based probes could be a promising and inexpensive component of a
metagenomic biosurveillance pipeline.

\section*{Acknowledgements}

The authors would like to acknowledge the use of AI-assisted tools during the preparation of this work. ChatGPT and Claude were used to support drafting, writing, and refinement of the manuscript, while Claude Code was used to assist with code development and implementation.

\rev{This work was conducted as part of the AIxBio Hackathon 2026, hosted by BlueDot Impact, Apart Research, and the Cambridge Biosecurity Hub. We thank the organizers for providing the environment and collaborative setting that enabled this project.}

\section*{CRediT Author Statement}

\textbf{Jeremy Guntoro:} Software; Investigation; Writing – Original Draft; Writing – Review \& Editing.

\textbf{Alexander Dack:} Conceptualization; Software; Investigation; Writing – Original Draft; Writing – Review \& Editing.

\textbf{Dylan Danno:} Software; Investigation; Writing – Original Draft.

\textbf{Michaela Jančovičová:} Data Curation; Investigation; Visualization.

\textbf{Križan Jurinović:} Data Curation; Investigation; Visualization.

\textbf{Vanessa Smilansky:} Data Curation; Investigation; Visualization.

\section*{Code and Data Availability Statement}

For biosecurity and dual-use risk mitigation reasons, the code and data associated with this work are not publicly released at this time, and the repository is kept private. However, we are committed to responsible scientific transparency and reproducibility. Access to the codebase and relevant data can be provided upon reasonable request to appropriate individuals, subject to review to ensure compliance with biosecurity and ethical considerations.

\bibliography{example_paper}
\bibliographystyle{icml2025}

%%%%%%%%%%%%%%%%%%%%%%%%%%%%%%%%%%%%%%%%%%%%%%%%%%%%%%%%%%%%%%%%%%%%%%%%%%%%%%%
% APPENDIX
%%%%%%%%%%%%%%%%%%%%%%%%%%%%%%%%%%%%%%%%%%%%%%%%%%%%%%%%%%%%%%%%%%%%%%%%%%%%%%%
\newpage
\appendix
\crefalias{section}{appendix}
\onecolumn

% Appendix float numbering: restart and prefix with the appendix letter (A1, A2, ...).
\setcounter{table}{0}
\setcounter{figure}{0}
\renewcommand{\thetable}{\Alph{section}\arabic{table}}
\renewcommand{\thefigure}{\Alph{section}\arabic{figure}}

\section{Implementation details}
\label{app:implementation}

\paragraph{Infrastructure.}
All activation extraction and probe training were run on Modal, a serverless cloud platform. Activation extraction was performed on NVIDIA H100 GPUs; probe training and evaluation were run on CPU.

\paragraph{Dataset splits.}
\Cref{tab:splits} gives the train/validation/test record counts for each probe.
The AMR, VFDB, and SynGenome probes use 1:1 positive/negative balancing, so each count in Table A1 splits evenly into positive and negative records. \rev{The SynGenome task is split at the individual-record level and predicts AMR-associated prompt labels.} The per-drug-class AMR split is shared across the five drug-class probes and is taken over AMR CDS from the five most common drug classes only (1,227 train / 275 val / 264 test); for each probe this pool is partitioned one-vs-rest into positives (that class) and negatives (the other four), as given in Table A2. The NEG, STRESS, and VIRULENCE transfer-negative pools are held separately and are used only at test time (Table 3). The SynGenome counts are for the balanced in-distribution set on which the \rev{prompt-label results} are reported, which is distinct from the larger unbalanced SynGenome pool used only for out-of-distribution checks.

\begin{table}[htbp]
  \centering
  \caption{Per-probe train/validation/test record counts. The split is balanced
    1:1, so each entry is (positive $+$ negative). The short-read set is
    evaluation-only: the trained AMR binary probe is applied to simulated reads,
    so it has no training or validation split; its decision threshold is selected on 301 bp segments of the validation CDS. Per-drug-class AMR splits are given
    separately in \cref{tab:splits-perclass}.}
  \label{tab:splits}
  \begin{tabular}{l r r r}
    \toprule
    Probe & Train & Val & Test \\
    \midrule
    AMR binary / attention & 3{,}122 & 664 & 672 \\
    VFDB bacterial virulence         & 11{,}922 & 2{,}392 & 336 \\
    \rev{SynGenome prompt-label task} & 3{,}122 & 664 & 672 \\
    AMR short reads        & ---     & ---  & 1{,}340 \\
    \bottomrule
  \end{tabular}
\end{table}
\FloatBarrier

\begin{table}[h]
    \centering
    \caption{Per-drug-class AMR dataset composition. The five probes are trained and
      evaluated one-vs-rest on AMR CDS from the five most common drug classes. For each class, positives are that class's CDS and negatives
      are the other four classes' CDS in the same split, so positives and negatives
      partition the same pool and each row sums to the column total (1{,}227 train /
      275 val / 264 test). AMR CDS from the remaining, less common drug classes are
      not used by these probes. At test time each probe is additionally evaluated
      against three shared transfer-negative pools drawn from outside the AMR
      classes: NEG $=414$, STRESS $=47$, and VIRULENCE $=31$ records. NEG, STRESS
      and VIRULENCE records are never used in training.}
    \label{tab:splits-perclass}
    \begin{tabular}{l c c c}
      \toprule
      Drug class & Train (pos / neg) & Val (pos / neg) & Test (pos / neg) \\
      \midrule
      Glycopeptide        & 389 / 838    & 92 / 183 & 81 / 183 \\
      Macrolide           & 396 / 831    & 70 / 205 & 79 / 185 \\
      Beta-lactam         & 279 / 948    & 71 / 204 & 61 / 203 \\
      Aminoglycoside      & 83 / 1{,}144 & 27 / 248 & 21 / 243 \\
      Phenicol            & 80 / 1{,}147 & 15 / 260 & 22 / 242 \\
      \midrule
      Pool & 1{,}227      & 275      & 264 \\
      \bottomrule
    \end{tabular}
  \end{table}

\paragraph{Negative matching hierarchy.}
Matched negatives for the AMR and VFDB datasets were selected by the ordered
fallback hierarchy in \cref{tab:negmatch}, taking the first tier that yields a
candidate for each positive; a positive with no match at any tier is emitted
unpaired.

\begin{table}[htbp]
  \centering
  \caption{Ordered negative-matching fallback hierarchy for the AMR and VFDB
    datasets. Tiers are tried in order; the first to yield a candidate is used.
    The AMR and VFDB columns give the number of positives matched (or unmatched) at each tier
    (of 2{,}230 and 7{,}370 respectively).}
  \label{tab:negmatch}
  \begin{tabular}{l l l l r r}
    \toprule
    Tier & Mobilome & Length & GC & AMR & VFDB \\
    \midrule
    1               & excluded & $\pm$20\% & $\pm$0.05 & 2{,}222 & 6{,}982 \\
    2               & allowed  & $\pm$20\% & $\pm$0.05 & 2       & 118 \\
    3               & allowed  & $\pm$20\% & $\pm$0.10 & 5       & 105 \\
    4               & allowed  & $\pm$50\% & $\pm$0.10 & 0       & 120 \\
    none (unpaired) & ---      & ---       & ---       & 1       & 45 \\
    \bottomrule
  \end{tabular}
\end{table}

\paragraph{Probe architectures.}
The linear probe is implemented as \texttt{nn.Linear(4096, 1)}. Per-token logits $\{z_i\}_{i=1}^{n}$ are aggregated to a per-region score by mean-pooling $s_{\text{mean}} = \frac{1}{n}\sum_i z_i$. The attention probe has three learned parameters: a query and a weight vector (each \texttt{nn.Parameter} of size 4096) and a scalar bias. The query pools the per-token activations into a single vector by softmax attention ($\alpha = \mathrm{softmax}(\langle q, h_i\rangle/\sqrt{d})$); the weight and bias then map that pooled vector to a scalar logit. 

\paragraph{Training configuration.}
Probes are trained with PyTorch's \texttt{BCEWithLogitsLoss}, which applies the sigmoid and binary cross-entropy in a single numerically stable step directly on the per-token logits. For the AMR binary probe the positive class is a minority ($\sim 12\%$ of tokens across positive and negative regions), so positive tokens are upweighted via \texttt{pos\_weight}\,$=20.0$; the CDS-only datasets, which are balanced at the
region level, use \texttt{pos\_weight}\,$=1.0$.

All probes were trained with the Adam optimiser at $\text{lr}=10^{-3}$ and
seed $42$. The \rev{SynGenome prompt-label probe} was trained on 2 epochs, while the per-drug-class AMR probe was trained on 5 epochs per class. All other probes were trained on 5 epochs. 

\end{document}